\documentclass[12pt]{iopart}
\eqnobysec

\usepackage{iopams,graphicx,amssymb}

\def\eps{\varepsilon}
\def\Dm{\widetilde{\cal D}_{\mu}}

\def\D{{\cal D}}

\def\S{{\cal S}}

\def\p{{\bf p}}

\def\bfr{{\bf r}}
\def\bfx{{\bf x}}
\def\bfv{{\bf v}}

\begin{document}

\title[Effects of turbulent mixing on the nonequilibrium critical behaviour]
{Effects of turbulent mixing on the nonequilibrium critical behaviour}

\author{N V Antonov, V I Iglovikov and A S Kapustin}

\address{Department of Theoretical Physics, St.~Petersburg University,
Uljanovskaja 1, St.~Petersburg, Petrodvorez, 198504 Russia}

\ead{nikolai.antonov@pobox.spbu.ru}

\begin{abstract}
We study effects of turbulent mixing on the critical behaviour of a
nonequilibrium system near its second-order phase transition between
the absorbing and fluctuating states. The model describes the spreading
of an agent (e.g., infectious disease) in a reaction-diffusion system
and belongs to the universality class of the directed bond percolation
process, also known as simple epidemic process, and is equivalent to
the Reggeon field theory. The turbulent advecting velocity field is
modelled by the Obukhov--Kraichnan's rapid-change ensemble: Gaussian
statistics with the correlation function
$\langle vv\rangle \propto \delta(t-t') \, k^{-d-\xi}$, where $k$ is
the wave number and $0<\xi<2$ is a free parameter. Using the field
theoretic renormalization group we show that, depending on the relation
between the exponent $\xi$ and the spatial dimension $d$, the system
reveals different types of large-scale asymptotic behaviour, associated
with four possible fixed points of the renormalization group equations.
In addition to known regimes (ordinary diffusion, ordinary directed
percolation process, and passively advected scalar field), existence
of a new nonequilibrium universality class is established, and the
corresponding critical dimensions are calculated to first order of the
double expansion in $\xi$ and $\varepsilon=4-d$ (one-loop approximation).
It turns out, however, that the most realistic values $\xi=4/3$
(Kolmogorov's fully developed turbulence) and $d=2$ or 3 correspond to
the case of passive scalar field, when the nonlinearity of the Reggeon
model is irrelevant and the spreading of the agent is completely
determined by the turbulent transfer.
\end{abstract}


\maketitle

\section{Introduction} \label{sec:Intro}

Over the past few decades, constant interest has been attracted by the
spreading processes and corresponding nonequilibrium phase transitions;
see Refs. \cite{Hinr}--\cite{Beate} and the literature cited therein.
Spreading processes are ubiquitous in the Nature and are encountered in
physical, chemical, biological, ecological and sociological systems:
autocatalytic reactions, percolation in porous media, forest fires, epidemic
diseases, and so on. For definiteness, in the following we will use the
terminology of the latter case (spreading of a disease).

Depending on the conditions, the spreading of an agent (disease) can either
continue over the whole population, or terminate after some time. In the
first case, the system evolves to a stationary (but not thermally
equilibrium) active state, in which the sick and healthy individuals coexist,
passing repeatedly through infection and healing events, and their densities
are fluctuating (random) quantities. In the second case, when the probability
of being healed is large enough in comparison with the probability to be
infected, the system is ``trapped'' in an absorbing (inactive) state where
all the individuals are healthy and all the fluctuations cease completely.
The transitions between these fluctuating and absorbing phases are
continuous; they are especially interesting as examples of nonequilibrium
critical behaviour.

By analogy with the equilibrium second-order transitions \cite{Zinn,Book3}
it is expected that, near the critical point, many details of a specific
spreading process become irrelevant, so that the critical behaviour of
different systems appears identical and can be described by a certain
{\it universality class}. The aim of the theory is to identify possible
universality classes and to calculate their universal characteristics (such
as critical exponents, scaling functions) on the basis of an appropriate
theoretic model and within a controlled approximation or a regular
perturbation scheme.

As a rule, the spreading phenomena are modelled by various stochastic
reaction-diffusion processes on a lattice; see Ref. \cite{Hinr} for a
detailed discussion. In the continuum limit, they can be mapped onto
certain field theoretic models with the aid of special techniques
\cite{MSR,Doi}. Then the powerful tools \cite{Zinn,Book3} of the field
theoretic renormalization group (RG) can be applied to the investigation
of their critical behaviour.

The most typical processes belong to the
universality class of the so-called {\it directed bond percolation process}
\cite{Hinr}--\cite{JT}, which in the field theoretic formulation is
equivalent to the well-known Reggeon field theory \cite{Gribov}. It was
conjectured in \cite{Conj} that the critical behaviour always belongs to
this ``DP class'' provided the absorbing state of the process is unique,
the order parameter is one-dimensional, there are no specific symmetries,
no coupling with additional ``slow'' degrees of freedom and no long-range
interactions. Thus the DP process is expected \cite{Conj} to play in the
theory of noneqilibrium phase transitions the same paradigmatic role as
the standard $\lambda \phi^{4}$ model \cite{Zinn,Book3} does in the theory
of equilibrium critical behaviour. It is also sometimes referred to as
simple epidemic process with recovery or as Gribov's process; the
stochastic version of the Schl\"{o}gl's first reaction also belongs to this
class \cite{Hinr}--\cite{JT}. The corresponding critical behaviour is rather
well understood: above the upper critical dimension $d_{c}=4$ the critical
exponents are given by the mean-field theory, and below $d_{c}$ they have
been calculated to second order of the expansion in the deviation
$\eps=d_{c}-d$ of the spatial dimension $d$ from the upper critical value;
see the discussion and references in Ref.~\cite{JT}. Recently, the first
experimental observation of this universality class was achieved in
the transition between two topologically different turbulent states of
electrohydrodynamic convection of a nematic liquid crystal \cite{Liquid},
with a firm agreement between theoretical and experimental values of
all the critical exponents.

Extensive numerical and analytical investigations of various models of the
spreading processes have shown that effects of immunization
\cite{Hinr}--\cite{JT}, \cite{Levy}, long-range interactions
\cite{LevyT}--\cite{Levy2}, presence of quenched (static) disorder
\cite{Quench} or critical fluctuations \cite{Beate} in the medium where
the spreading occurs can lead to appearance of completely new universality
classes, with different values of the upper critical dimensions $d_{c}$,
different sets of critical exponents and interesting crossovers between
various critical regimes.

In this paper we will consider the spreading of a nonconserved agent in
a turbulent medium and study the effects of turbulent stirring and mixing
on the critical behaviour near the phase transition between the absorbing
and the fluctuating phases. Turbulence is hardly avoidable in the chemical
catalytic reactions and can play decisive role if a disease is transferred
by flying insects or birds.

Much work has been devoted to the investigation of the effects of various
kinds of imposed deterministic or chaotic flows (laminar shear flows,
turbulent convection and so on) on the behaviour of the critical or nearly
critical fluids, like binary liquid mixtures near and below their consolution
points; see e.g. the papers \cite{Onuki}--\cite{Chaotic} and references
therein. It was shown that the flow can destroy the usual critical behaviour,
typical of the $\lambda \phi^{4}$ model, which changes to the mean-field
behaviour \cite{Onuki,Beysens} or to a complex behaviour described by new
nonequilibrium universality classes \cite{Satten,AHH,Alexa}. This problem
is closely related to another interesting issue: the effects of imposed
flows on the dynamics of phase ordering -- the growth of order through
domain coarsening (spinodal decomposition) and formation of various
nontrivial spatiotemporal patterns; see \cite{Shear2}--\cite{Chaotic}.

Theoretical description of fully developed turbulence on the basis of a
microscopic dynamic model (like e.g. the stirred Navier-Stokes equation)
remains essentially an open problem \cite{Legacy}. In this paper, we
will describe the turbulent mixing by a Gaussian velocity ensemble
with prescribed statistics, which dates back to Obukhov, Batchelor,
Kazantsev and Kraichnan and now is widely known as the rapid-change model
\cite{FGV}. In spite of their relative simplicity, the models of passive
(no feedback on the velocity) scalar fields advected by such ``synthetic''
velocity ensembles have attracted serious attention recently because of the
deep insight they offer into the origin of intermittency and anomalous
(multi)scaling of the genuine turbulent heat or mass transport; see
the review paper \cite{FGV} and references therein. The RG approach
to that problem is reviewed in Ref. \cite{JphysA}.

In the rapid-change model, the pair velocity correlation function is taken
in the form $\langle vv\rangle \propto \delta(t-t') \, k^{-d-\xi}$, where
$k$ is the wave number and $0<\xi<2$ a free parameter with the
most realistic (Kolmogorov) value $\xi=4/3$. Vanishing of the correlation
time ensures the Galilean symmetry of the problem, while a power-law
dependence on the wave number $k$ mimics the real scaling properties
of fully developed turbulence.

The plan of the paper is as follows. In section~\ref{sec:QFT} we present
the detailed description of the model and its field theoretic formulation.
In section~\ref{sec:Reno} we analyze the ultraviolet (UV) divergences of
the model, relaying upon the canonical dimensions and additional
symmetry considerations. We show that the model is multiplicatively
renormalizable and write the renormalized action functional.

In section~\ref{sec:RGE} we derive the RG equations and introduce the RG
functions ($\beta$ functions and anomalous dimensions $\gamma$).
In section~\ref{sec:FPS} we identify four possible infrared (IR) attractive
fixed points of the RG equations, and identify their ranges of stability
in the $\eps$--$\xi$ plane.

These fixed points correspond to four possible critical regimes of the
model, with different sets of critical dimensions, as discussed in
section~\ref{sec:DimeNS}.
Three of them correspond to already known regimes: free (Gaussian)
field theory, linear passive scalar advection by the Obukhov--Kraichnan
ensemble (the nonlinearity in the agent's density appears irrelevant),
and the ordinary Gribov process (mixing by the turbulent field is
irrelevant). The fourth fixed point corresponds to a new nonequilibrium
universality class, with the new set of critical exponents that depend
on the both parameters $\eps$ and $\xi$ and can be systematically
calculated as double series in these parameters.

The practical calculation of the renormalization constants, RG functions,
regions of stability and critical dimensions is accomplished to the
leading-order (one-loop) approximation; some of the results, however,
are exact (valid to all orders of the double $\eps$--$\xi$ expansion).

In section~\ref{sec:DimeNS} we also consider, as a special consequence
of the general scaling relations, the temporal evolution of a cloud of the
advected agent's particles (infected individuals), which differs,
in general, from the well-known ``1/2 law'' for ordinary diffusion.
Section~\ref{sec:Conc} is reserved for discussion and conclusion.

\section{Description of the model. Field theoretic formulation}
\label{sec:QFT}

In the Lagrangian formulation, spreading of an agent is described by a
continuous stochastic diffusion-reaction equation of the form
\begin{equation}
\partial_{t} \psi(t,{\bf x}) = \lambda_{0} \left\{ (-\tau_{0} +
\partial^{2}) \psi(t,{\bf x}) - g_{0} \psi^{2}(t,{\bf x})/2 \right\}
+ \zeta (t,{\bf x}),
\label{stoh}
\end{equation}
where $\psi(t,{\bf x})>0$ is the agent's density, $\partial^{2}$ is the
Laplace operator, $\lambda_{0}$ and $g_{0}$ are positive parameters,
$\tau_{0}$ --- deviation of the infection probability
from its critical value (analog of $\tau_{0}\propto (T-T_{c})$ in
equilibrium systems).
The random Gaussian noise $\zeta(t,{\bf x})$
with zero mean and a given correlation function
\begin{equation}
\langle \zeta (t,{\bf x})\zeta (t',{\bf x'}) \rangle = g_{0}\lambda_{0}
\psi(t,{\bf x}) \delta(t-t')\delta^{(d)}({\bf x}-{\bf x}')
\label{shum}
\end{equation}
mimics fluctuations in the system; $d$ is the dimension of the
${\bf x}$ space.

According to the general theorem \cite{MSR} (see also the monographs
\cite{Zinn,Book3}), stochastic problem (\ref{stoh}), (\ref{shum})
is equivalent to the field theoretic model of the doubled set of
fields with action functional
\begin{eqnarray}
\S(\psi,\psi^{\dag}) =  \psi^{\dag}
(-\partial_{t}+\lambda_{0} \partial^{2}- \lambda_{0}\tau_{0}) \psi
+ \frac{g_{0}\lambda_{0}}{2} \left( (\psi^{\dagger})^2\psi -
\psi^{\dagger}\psi^2  \right).
\label{action}
\end{eqnarray}
Here, $\psi^{\dag}=\psi^{\dag}(t,{\bf x})$ is the auxiliary
``response field'' and the integration over the arguments
of the fields is implied, for example
\[  \psi^{\dag}\partial_{t}\psi = \int dt \int d{\bf x}
\psi^{\dag}(t,{\bf x})\partial_{t}\psi(t,{\bf x}). \]
This means that statistical averages of random quantities in the original
stochastic problem (\ref{stoh}), (\ref{shum}) can be represented as
functional averages over the full set of fields with the weight
$\exp {\cal S}(\Phi)$, and can therefore be viewed as the Green
functions of the field theoretic model with the action (\ref{action}).
In particular, the linear response function of
the problem (\ref{stoh}), (\ref{shum}) is given by the Green function
$G=\langle \psi^{\dag} \psi \rangle$ of the model (\ref{action}).

More rigorous derivation of the field theory is based on the
so-called master equation for the original reaction-diffusion process
on the lattice, which involves a microscopic Hamiltonian written in terms
of the creation-annihilation operators with further representation of the
corresponding evolution operator by the coherent-state functional integral
\cite{Doi}. For the case at hand the resulting action functional
coincides with (\ref{action}) up to irrelevant terms; see also discussion
of that approach in Refs. \cite{Tauber,JT}.

One can argue that in the
perturbation theory the condition $\psi>0$ can be neglected \cite{Conj};
then the model (\ref{action}) becomes equivalent to the Reggeon field theory
\cite{Gribov} and acquires the symmetry with respect to the transformation
\begin{equation}
\psi(t,{\bf x})\to \psi^{\dagger}(-t,-{\bf x}),
\quad \psi^{\dagger}(t,{\bf x})\to \psi(-t,-{\bf x}),
\quad g_{0}\to -g_{0}.
\label{symm}
\end{equation}
Reflection of the constant $g_{0}$ is in fact unimportant because, as can
easily be seen, the actual expansion parameter in the perturbation theory
is $g_{0}^{2}$ rather than $g_{0}$ itself.

The model (\ref{action}) corresponds to a standard Feynman diagrammatic
technique with the only bare propagator
$G_{0}=\langle \psi \psi^{\dag} \rangle_{0}$ and the two triple vertices
$\sim (\psi^{\dagger})^2\psi$, $\psi^{\dagger}\psi^2$
In the time-momentum and frequency-momentum representation $G_{0}$
has the forms
\begin{equation}
G_{0}(t,k)=\theta(t) \exp \left\{ - \lambda_{0}
(k^{2}+\tau_{0}) \right\} \leftrightarrow
G_{0}(\omega,k)= \frac{1}{-{\rm i}\omega+\lambda_{0}
\left(k^{2}+\tau_{0}\right)}.
\label{lines}
\end{equation}
Here $\theta(\dots)$ is the Heaviside step function, so that the propagator
(\ref{lines}) is retarded. Then from the analysis of the diagrams one can
check that the Green functions built solely from the field $\psi$
or solely from $\psi^{\dag}$ necessarily contain closed circuits of
retarded propagators (\ref{lines}) and therefore vanish identically.
For the functions $\langle \psi^{\dag} \dots \psi^{\dag} \rangle$ this
fact is a general consequence of the causality, which is valid
for any stochastic model; see e.g. the discussion in \cite{Book3}.
Then vanishing of the functions $\langle \psi \dots \psi \rangle$
can be viewed as a consequence of the symmetry (\ref{symm}).

The stability in dynamical models like (\ref{action}) implies that all the
small fluctuations are damped out, so that the exact response function
$G=\langle \psi^{\dag} \psi \rangle$ must decay for $t\to\infty$;
see e.g. sec. 5.5 in \cite{Book3}. Then from the expression (\ref{lines}),
which is the zero order approximation for $G$, we conclude that the
stability of the perturbative stationary state is lost for $\tau_{0}=0$,
and for $\tau_{0}<0$ the perturbations with smallest momenta grow in time.
This growth is stabilized by the appearance of the nonzero constant mean
$\langle \psi \rangle$ and the higher-order correlation functions of the
agent field $\psi$, so that the symmetry (\ref{symm}) is spontaneously
broken. This is exactly the phase transition from the absorbing (normal)
to the fluctuating (anomalous) states.

Coupling with the velocity field ${\bf v}= \{ v_{i}(t,\bfx) \}$
is introduced by the replacement
\begin{eqnarray}
\partial_{t} \to \nabla_{t} = \partial_{t} + v_{i} \partial_{i},
\label{nabla}
\end{eqnarray}
in (\ref{stoh}) and (\ref{action}), where
$\partial_i = \partial /\partial x_{i}$ and $\nabla_{t}$ is the
Lagrangian (Galilean covariant) derivative. We will consider the case
of incompressible flow, then the velocity field is divergence-free
(transverse): $\partial _i v_{i}=0$. In the real problem, the field
${\bf v}(t,{\bf x})$ satisfies the Navier--Stokes equation. We will employ
the rapid-change model, where the velocity obeys a Gaussian distribution
with zero mean and correlation function
\begin{eqnarray}
\langle v_{i}(t, \bfx) v_{j}(t',{\bf x'})\rangle =  \delta(t-t')\,
D_{ij}(\bfr), \quad \bfr = \bfx-{\bf x'}
\label{white}  \\
D_{ij}(\bfr) = D_{0}\, \int_{k>m} \frac{d{\bf k}}{(2\pi)^{d}} \,
P_{ij}({\bf k})\, \frac{1}{k^{d+\xi}}\, \exp ({\rm i} {\bf kr} ),
\quad k\equiv |{\bf k}|
\label{Kraich}
\end{eqnarray}
where $P_{ij}({\bf k}) = \delta_{ij} - k_i k_j / k^2$ is the transverse
projector, $D_{0}>0$ is an amplitude factor and $0<\xi<2$ is a free
parameter with the most realistic (``Kolmogorov'') value $\xi=4/3$.
The cutoff in the integral (\ref{Kraich}) from below at $k=m$, where
$m\equiv 1/{\cal L}$ is the reciprocal of the integral turbulence scale
${\cal L}$, provides the IR regularization. Its precise form is
unimportant; the sharp cutoff is the simplest choice for the practical
calculations.

The full problem is equivalent to the field theoretic model of the three
fields $\Phi = \{ \psi, \psi^{\dag}, \bfv \} $ with the action functional
\begin{eqnarray}
{\cal S}(\Phi) &=& \psi^{\dag}
(-\nabla_{t}+\lambda_{0} \partial^{2}- \lambda_{0}\tau_{0}) \psi +
\frac{\lambda_{0} g_{0}}{2} \left( (\psi^{\dagger})^2\psi -
\psi^{\dagger}\psi^2  \right) + \S(\bfv),
\label{Action}
\end{eqnarray}
which is obtained from (\ref{action}) by the replacement (\ref{nabla})
and adding the term corresponding to the Gaussian averaging over the field
$\bfv$ with the correlator (\ref{Kraich}):
\begin{eqnarray}
\S(\bfv) = -\frac{1}{2} \int dt \int d{\bf x} \int d{\bf x'}
v_{i} (t,\bfx) D_{ij}^{-1}(\bfr) v_{j} (t, {\bf x'}),
\label{Sv}
\end{eqnarray}
where
\[ D_{ij}^{-1}(\bfr) \propto D_{0}^{-1} r^{-2d-\xi} \]
is the kernel of the inverse linear operation for the function
$D_{ij}(\bfr)$ in (\ref{Kraich}).

In addition to (\ref{lines}), the Feynman diagrams for the model
(\ref{Action}) involve the propagator $\langle vv \rangle_{0}$ specified
in (\ref{white}), (\ref{Kraich}) and the new vertex
$-\psi^{\dag} (v\partial) \psi $.

The role of the coupling constants in the ordinary perturbation theory
is played by the two parameters
\begin{equation}
u_{0} = g^{2}_{0} \sim \Lambda^{4-d}  , \qquad
w_{0} = D_{0}/\lambda_{0} \sim \Lambda^{\xi}.
\label{charges}
\end{equation}
The last relations, following from the dimensionality considerations,
(more precisely, see the next section) define the typical UV momentum
scale $\Lambda$. By rescaling the fields, the constant $w_{0}$ can be
placed in front of the interaction term $-\psi^{\dag} (v\partial) \psi $,
which is more familiar for the field theory. We do not do it, however,
in order not to spoil the natural form of the covariant derivative, and
thus assign the factor $w_{0}$ to the propagator $\langle vv \rangle_{0}$.

\section{Canonical dimensions, UV divergences and the renormalization}
\label{sec:Reno}

It is well known that the analysis of UV divergences is based on the analysis
of canonical dimensions (``power counting''); see e.g. \cite{Zinn,Book3}.
Dynamic models of the type (\ref{Action}), in contrast to static ones, have
two independent scales: the time scale $T$ and the length scale $L$. Thus
the canonical dimension of some quantity $F$ (a field or a parameter in
the action functional) is completely characterized by two numbers, the
frequency dimension $d_{F}^{\omega}$ and the momentum dimension $d_{F}^{k}$,
defined such that $[F] \sim [T]^{-d_{F}^{\omega}} [L]^{-d_{F}^{k}}$. These
dimensions are found from the obvious normalization conditions
\[ d_k^k=-d_{\bf x}^k=1,\ d_k^{\omega} =d_{\bf x}^{\omega }=0,\
d_{\omega }^k=d_t^k=0, \ d_{\omega }^{\omega }=-d_t^{\omega }=1, \]
and from the requirement
that each term of the action functional be dimensionless (with
respect to the momentum and frequency dimensions separately).
Then, based on $d_{F}^{k}$ and $d_{F}^{\omega}$,
one can introduce the total canonical dimension
$d_{F}=d_{F}^{k}+2d_{F}^{\omega}$ (in the free theory,
$\partial_{t}\propto\partial^{2}$), which plays in the theory of
renormalization of dynamical models the same role as
the conventional (momentum) dimension does in static problems;
see Chap.~5 of \cite{Book3}.

The dimensions for the model (\ref{Action}) are given in table~\ref{table1},
including renormalized parameters (without subscript ``o''), which will be
introduced later on.
From table~\ref{table1} or, equivalently, from the relations (\ref{charges})
it follows that the model is logarithmic (the both coupling constants
$g_{0}$ and $w_{0}$ are simultaneously dimensionless) at $d=4$ and $\xi=0$.
Thus the UV divergences in the Green functions manifest themselves as
poles in $\eps = 4-d$, $\xi$ and, in general, their linear combinations.
\begin{table}
\caption{Canonical dimensions of the fields and parameters in the
model (\protect\ref{Action}).}
\label{table1}
\begin{tabular}{cccccccccc}
\br
$F$ & $\psi$, $\psi^{\dag}$ & $ {\bf v} $ &  $\lambda$, $\lambda_{0}$ &
$\tau$, $\tau_{0}$ &  $m,\mu, \Lambda $ & $D_{0}$ & $u_{0}=g_{0}^{2}$ &
$w_{0}$ & $g$, $u$, $w$ \\
\br
%\tableline
$d_{F}^{k}$ & $d/2$ & $-1$ & $-2$  & 2 & 1 & $-2+\xi$ & $4-d$ & $\xi$ & 0 \\
\mr
$d_{F}^{\omega }$ & 0 & 1 & 1 & 0 & 0 & 1 & 0 & 0 & 0 \\
\mr
$d_{F}$ & $d/2$ & 1 & 0 & 2 & 1 & $\xi$ & $4-d$  & $\xi$ & 0 \\
\br
\end{tabular}
\end{table}

The total canonical dimension of an arbitrary 1-irreducible Green function
$\Gamma = \langle\Phi \cdots \Phi \rangle _{\rm 1-ir}$
is given by the relation \cite{Book3}
\begin{equation}
d_{\Gamma }=d_{\Gamma }^k+2d_{\Gamma }^{\omega }=
d+2-N_{\Phi }d_{\Phi},
\label{dGamma}
\end{equation}
where $N_{\Phi}=\{N_{\psi},N_{\psi^{\dag}}, N_{v}\}$ are the numbers of
corresponding fields entering into the function $\Gamma$, and the summation
over all types of the fields is implied.

The total dimension $d_{\Gamma}$ in logarithmic theory (that is, at
$\eps=\xi=0$) is the formal index of the UV divergence
$\delta_{\Gamma}=d_{\Gamma}|_{\eps=\xi=0}$.
Superficial UV divergences, whose removal requires counterterms, can be
present only in those functions $\Gamma$ for which $\delta_{\Gamma}$ is
a non-negative integer.
From table~\ref{table1} and (\ref{dGamma}) we find
\begin{equation}
\delta_{\Gamma}= 6 - 2N_{\psi} - 2N_{\psi^{\dag}} - N_{v}.
\label{IndeX}
\end{equation}
We recall that in our model nonvanishing Green functions must involve the
both fields $\psi$ and $\psi^{\dag}$ (see the discussion in the preceding
section), so in (\ref{IndeX}) it is sufficient to take
$N_{\psi} \ge 1$ and (simultaneously) $N_{\psi}^{\dag} \ge 1$.
Straightforward analysis of the expression
(\ref{IndeX}) then shows that superficial UV divergences can be present
only in the following 1-irreducible functions:
\[ \langle \psi^{\dag} \psi \rangle \quad (\delta=2) \quad
{\rm with\ the\ counterterms} \quad \psi^{\dag}\partial_{t}\psi, \
\psi^{\dag}\partial^{2}\psi, \ \psi^{\dag}\psi, \]
\[ \langle \psi^{\dag} \psi\psi \rangle \quad (\delta=0) \quad
{\rm with\ the\ counterterm} \quad \psi^{\dag} \psi^{2}, \]
\[ \langle \psi^{\dag} \psi^{\dag}\psi \rangle \quad (\delta=0) \quad
{\rm with\ the\ counterterm} \quad (\psi^{\dag})^{2} \psi, \]
\[ \langle \psi^{\dag} \psi v \rangle \quad (\delta=1), \]
for which the counterterm necessarily reduces to the form
$\psi^{\dag} (v\partial) \psi = - \psi (v\partial) \psi^{\dag}$
owing to the transversality of the velocity field. All such terms are
present in the action (\ref{Action}),
so that our model appears multiplicatively renormalizable.

The superficial divergence in the function
$\langle \psi^{\dag} \psi vv \rangle$ with $\delta=0$ and the
counterterm $\psi^{\dag} \psi v^{2}$,
allowed by the dimension, is in fact forbidden by the Galilean symmetry.
Furthermore, the latter requires that the counterterms
$\psi^{\dag}\partial_{t}\psi$ and $\psi^{\dag} (v\partial) \psi $
enter the renormalized action only
in the form of the Lagrangian derivative $\psi^{\dag}\nabla_{t}\psi$.

Strictly speaking, the arguments based on the Galilean symmetry are
applicable only to the velocity field governed by the Navier--Stokes
equation, and generally become invalid for synthetic Gaussian velocity
ensembles. It turns out, however, that for a Gaussian ensemble of the type
(\ref{white}) with {\it vanishing} correlation time the Galilean symmetry
of the counterterms indeed takes place; see e.g. \cite{FGV}. This issue,
along with the consequences of the Galilean invariance for the
renormalization, is discussed in the appendix A of Ref. \cite{Alexa}
in detail. The proof given there is fully applicable to the model
(\ref{Action}). From the symmetry (\ref{symm}) it also follows that the
trilinear counterterms enter the renormalized action as the single
combination $(\psi^{\dagger})^2\psi - \psi^{\dagger}\psi^2 $.

We thus conclude that the renormalized action can be written in the form
\begin{eqnarray}
{\cal S}_{R}(\Phi) &=&  \psi^{\dag} \left(- Z_{1} \nabla_{t}
+ Z_{2}\lambda \partial^{2}- Z_{3} \lambda\tau\right)
\psi + \nonumber \\ &+&
Z_{4} \frac{\lambda  g}{2} \left( (\psi^{\dagger})^2\psi -
\psi^{\dagger}\psi^2  \right)  + \S(\bfv).
\label{RenAct}
\end{eqnarray}
Here $\lambda$, $\tau$ and $g$ are renormalized analogs of the bare
parameters (with the subscripts ``o'') and $\mu$ is the reference mass
scale (additional arbitrary parameter of the renormalized theory).
Since the last term $\S(\bfv)$ given by (\ref{Sv}) is not renormalized,
the amplitude $D_{0}$ is expressed in renormalized parameters as
\begin{eqnarray}
D_{0} = w_{0} \lambda_{0}  = w\lambda \mu^{\xi}.
\label{RenD}
\end{eqnarray}

Expression (\ref{RenAct}) can be obtained by the multiplicative
renormalization of the fields $\psi \to \psi Z_{\psi}$,
$\psi^{\dag} \to \psi^{\dag} Z_{\psi^{\dag}}$,  $v \to vZ_{v}$
and the parameters:
\begin{eqnarray}
\lambda_{0} = \lambda Z_{\lambda}, \quad
\tau_{0} = \tau Z_{\tau},          \quad
g_{0} = g \mu^{\eps/2} Z_{g},      \quad
w_{0} = w \mu^{\xi} Z_{w} .
\label{Multy}
\end{eqnarray}
The renormalization constants in Eqs. (\ref{RenAct}) and (\ref{Multy})
are related as follows:
\begin{eqnarray}
Z_{1} = Z_{\psi} Z_{\psi^{\dagger}} = Z_{v}  Z_{\psi} Z_{\psi^{\dagger}}
\quad
Z_{2} = Z_{\psi} Z_{\psi^{\dagger}} Z_{\lambda}, \quad
Z_{3} = Z_{\psi} Z_{\psi^{\dagger}}Z_{\lambda}Z_{\tau},
\nonumber \\
Z_{4} = Z_{\psi} Z_{\psi^{\dagger}}^{2} Z_{g} Z_{\lambda} =
        Z_{\psi}^{2} Z_{\psi^{\dagger}} Z_{g} Z_{\lambda}, \quad
1= Z_{w} Z_{\lambda} .
\label{ZZ}
\end{eqnarray}
Resolving these relations with respect to the renormalization constants
of the fields and parameters gives
\begin{eqnarray}
Z_{v}=1, \quad Z_{\psi}= Z_{\psi^{\dagger}}=Z_{1}^{1/2}, \quad
Z_{\tau}= Z_{3}Z_{2}^{-1}, \nonumber \\
Z_{\lambda}= Z_{w}^{-1}= Z_{2}Z_{1}^{-1}, \quad
Z_{g} = Z_{4}Z_{2}^{-1}Z_{1}^{-1/2},
\label{Reso}
\end{eqnarray}
where the first equality is a consequence of the Galilean symmetry and the
second -- a consequence of the symmetry (\ref{symm}). The first relation
in the second line is a consequence of the absence of renormalization
of the term $\S (\bfv)$ in (\ref{RenAct}). For the coupling constant
$u_{0}=g^{2}_{0}$ introduced in (\ref{charges}) one has:
\begin{eqnarray}
u_{0} = u \mu^{\eps} Z_{u}, \quad  Z_{u}=Z_{g}^{2}.
\label{u}
\end{eqnarray}
The renormalization constants $Z_{1}$--$Z_{4}$ are calculated directly from
the diagrams, then the constants in (\ref{Multy}) are found from
(\ref{Reso}) and (\ref{u}).

The renormalization constants capture all the divergences at $\eps,\xi\to 0$,
so that the correlation functions of the renormalized model (\ref{RenAct})
have finite limits for $\eps$, $\xi = 0$ when expressed in renormalized
parameters $\lambda$, $\tau$ and so on. In practical calculations, we used
the minimal subtraction (MS) scheme, in which the renormalization constants
have the forms $Z_{i}=1+\,$ only singularities in $\eps$ and $\xi$, with the
coefficients depending on the two completely dimensionless parameters ---
renormalized coupling constants $u$ and $w$. To simplify the resulting
expressions, it is convenient to pass to the new couplings,
\begin{eqnarray}
u \to u/16\pi^2, \quad w \to w/16\pi^2;
\label{16p}
\end{eqnarray}
in what follows they will be denoted by the same symbols.

Explicit calculation in the first order in $u$ and $w$
(one-loop approximation) gives rather simple results:
\begin{eqnarray}
Z_{1} = 1 + \frac{u}{4\eps}, \quad
Z_{2} = 1 + \frac{u}{8\eps} - \frac{3w}{4\xi}, \quad   %\nonumber \\
Z_{3} = 1 + \frac{u}{2\eps}, \quad
Z_{4} = 1 + \frac{u}{\eps},
\label{Zone}
\end{eqnarray}
with the corrections of second order in $u$ and $w$ and higher. For $w=0$
one obtains (up to the notation) the well-known one-loop result for the
model (\ref{action}); cf. e.g. \cite{JT}), while for $u=0$ the exact result
for the rapid-change model is recovered; cf. e.g. \cite{JphysA}.
% Details of the calculation are given in appendix~A.

\section{RG functions and RG equations} \label{sec:RGE}

Let us recall an elementary derivation of the RG equations; detailed
exposition can be found in monographs \cite{Zinn,Book3}.
The RG equations are written for the renormalized Green functions
$G_{R} =\langle \Phi\cdots\Phi\rangle_{R}$, which differ from the original
(unrenormalized) ones $G =\langle \Phi\cdots\Phi\rangle$ only by
normalization (due to rescaling of the fields) and choice of
parameters, and therefore can equally be used for analyzing the critical
behaviour. The relation $\S_{R} (\Phi,e,\mu) = \S(\Phi,e_{0})$ between the
functionals (\ref{Action}) and (\ref{RenAct}) results in the relations
\begin{equation}
G(e_{0},\dots) = Z_{\psi}^{N_{\psi}} Z_{\psi^{\dagger}}^{N_{\psi^{\dagger}}}
G_{R}(e,\mu,\dots).
\label{multi}
\end{equation}
between the Green functions. Here, as usual, $N_{\psi}$ and
$N_{\psi^{\dagger}}$ are the numbers of corresponding fields
entering into $G$ (we recall that in our model $Z_{v}=1$);
$e_{0}=\{\lambda_{0}, \tau_{0}, u_{0}, w_{0} \}$ is the full set of
bare parameters and $e=\{ \lambda, \tau, u, w  \}$ are their renormalized
counterparts; the dots stand for the other arguments
(times/frequencies and coordinates/momenta).

We use $\widetilde{\cal D}_{\mu}$ to denote the differential operation
$\mu\partial_{\mu}$ for fixed $e_{0}$ and operate on both sides of the
equation (\ref{multi}) with it. This gives the basic RG differential
equation:
\begin{equation}
\left\{ {\cal D}_{RG} + N_{\psi} \gamma_{\psi} +
N_{\psi^{\dag}} \gamma_{\psi^{\dag}} \right\}
\,G_{R}(e,\mu,\dots) = 0,
\label{RG1}
\end{equation}
where ${\cal D}_{RG}$ is the operation $\widetilde{\cal D}_{\mu}$
expressed in the renormalized variables:
\begin{equation}
{\cal D}_{RG}\equiv {\cal D}_{\mu} + \beta_{u}\partial_{u} +
\beta_{w}\partial_{w}  -
\gamma_{\lambda}{\cal D}_{\lambda} - \gamma_{\tau}{\cal D}_{\tau}.
\label{RG2}
\end{equation}
Here we have written ${\cal D}_{x}\equiv x\partial_{x}$ for any variable
$x$, and the anomalous dimensions $\gamma$ are defined as
\begin{equation}
\gamma_{F}\equiv \Dm \ln Z_{F} \quad {\rm for\ any\ quantity} \ F,
\label{RGF1}
\end{equation}
and the $\beta$ functions for the two dimensionless couplings $u$ and $w$ are
\begin{equation}
\beta_{u} \equiv \widetilde {\cal D}_{\mu} u = u\,[-\eps-\gamma_{u}],
\quad
\beta_{w} \equiv \widetilde {\cal D}_{\mu} w = w\,[-\xi-\gamma_{w}],
\label{betagw}
\end{equation}
where the second equalities come from the definitions and the
relations (\ref{Multy}).

Equations (\ref{Reso}) result in the following relations between
the anomalous dimensions (\ref{RGF1}):
\begin{eqnarray}
\gamma_{\psi}= \gamma_{\psi^{\dagger}}= \gamma_{1}/2, \quad
\gamma_{\lambda} = -\gamma_{w}= \gamma_{2}-\gamma_{1}, \quad
\gamma_{v} =0 , \nonumber \\
\gamma_{\tau} = \gamma_{3}-\gamma_{2}, \quad
\gamma_{u} = 2\gamma_{4}-2\gamma_{2}- \gamma_{1} .
\label{gadf}
\end{eqnarray}

The anomalous dimension corresponding to a given renormalization constant
$Z_{F}$ is readily found from the relation
\begin{equation}
\gamma_{F} = \left(\beta_{u}\partial_{u}+\beta_{w}\partial_{w}\right)
\ln Z_{F} \simeq  - \left(\eps\D_{u}+\xi\D_{w}\right) \ln Z_{F}.
\label{GfZ}
\end{equation}
In the first relation, we used the definition (\ref{RGF1}), expression
(\ref{RG2}) for the operation $\Dm$ in renormalized variables, and the
fact that the $Z$'s depend only on the two completely dimensionless coupling
constants $u$ and $w$. In the second (approximate) relation, we retained only
the leading-order terms in the $\beta$ functions (\ref{betagw}), which is
sufficient for the first-order approximation. The factors $\eps$ and $\xi$
in (\ref{GfZ}) cancel the corresponding poles contained in the expressions
(\ref{Zone}) for the constants $Z_{F}$, which leads to the final UV finite
expressions for the anomalous dimensions. This gives:
\begin{eqnarray}
\gamma_{1} =  -\frac{u}{4}, \quad
\gamma_{2} = -\frac{u}{8}+\frac{3w}{4}, \quad
\gamma_{3} = - \frac{u}{2} , \quad
\gamma_{4} = -u,
\label{gammaOn2e}
\end{eqnarray}
and from (\ref{gadf}) one finally obtains:
\begin{eqnarray}
\gamma_{\psi}=\gamma_{\psi^{\dagger}}=-\frac{u}{8}, \quad
\gamma_{\lambda}=-\gamma_{w}= \frac{u}{8}+\frac{3w}{4} , \nonumber \\
\gamma_{\tau} = -\frac{3u}{8}-\frac{3w}{4},  \quad
\gamma_{u} =-\frac{3u}{2} -\frac{3w}{2},
\label{gammasfields}
\end{eqnarray}
with corrections of order $u^{2}$, $w^{2}$, $uw$ and higher.

\section{Fixed points and scaling regimes} \label{sec:FPS}

It is well known that the long-time large-distance asymptotic behaviour
of a renormalizable field theory is determined by the IR attractive fixed
points of the corresponding RG equations. In general, coordinates of the
possible fixed points are found from the requirement that all the $\beta$
functions vanish. In the model (\ref{Action}) the coordinates
$u_{*}$, $w_{*}$ are determined by the two equations
\begin{equation}
\beta_{u} (u_{*},w_{*}) = 0, \quad \beta_{w} (u_{*},w_{*})=0 ,
\label{points}
\end{equation}
with the $\beta$ functions given in (\ref{betagw}).
The type of a fixed point is determined by the matrix
\begin{equation}
\Omega=\{\Omega_{ij}=\partial\beta_{i}/\partial g_{j}\},
\label{OmegaDef}
\end{equation}
where $\beta_{i}$ is the full set of the $\beta$ functions and
$g_{j}= \{g,w\}$ is the full set of coupling constants. For an IR
attractive fixed point the matrix $\Omega$ is positive, that is,
the real parts of all its eigenvalues are positive.

From the definitions (\ref{betagw}) and explicit expressions
(\ref{gammasfields}) for the anomalous dimensions we derive the
following leading-order expressions for the $\beta$ functions:
\begin{equation}
\beta_{u} = u\, (-\varepsilon + 3u/2 +3w/2), \quad
\beta_{w} = w\, (-\xi +u/8+3w/4).
\label{betas2}
\end{equation}
From Eqs. (\ref{points}) and (\ref{betas2}) we can identify four different
fixed points. For the three of them, the matrix $\Omega$ appears triangular
and its eigenvalues are given by the diagonal elements $\Omega_{u} = \partial
\beta_{u} / \partial u$ and $\Omega_{w} = \partial \beta_{w} / \partial w$.

\

\noindent 1. Gaussian (free) fixed point:
$u_{*}=w_{*}=0$;  $\Omega_{u} = -\eps$,  $\Omega_{w} = -\xi$
(all these expressions are exact).

\

\noindent 2. $w_{*}=0$ (exact result to all orders), $u_{*}=2\eps/3$;
$\Omega_{u} = \eps$,  $\Omega_{w} = -\xi+\eps/12$.

\

In this regime, effects of the turbulent mixing are irrelevant in the
leading-order IR asymptotic behaviour; the basic critical exponents are
independent on $\xi$ and coincide to all orders with their counterparts
for the ``pure'' DP class \cite{Hinr}--\cite{JT}. However, the dependence
on $\xi$ appears in the {\it corrections} to the leading-order behaviour,
in particular, due to the correction exponent $\Omega_{w}$. Although the
expression for $\Omega_{u}$ is not exact (it has corrections of order
$\eps^{2}$ and higher), the inequality $\Omega_{u}>0$ is equivalent to
$\eps>0$ within the $\eps$ expansion.

\

\noindent 3. $u_{*}=0$, $w_{*}=4\xi/3$ (exact);
$\Omega_{u} = -\eps+2\xi$, $\Omega_{w} = \xi$ (exact).

\

In this regime, the nonlinearity
$(\psi^{\dag})^2\psi - \psi^{\dagger}\psi^2$
of the DP model is irrelevant, and we arrive at the rapid-change
model of a passively advected scalar field $\psi$ \cite{FGV}.
For that model, the $\beta$ function is given exactly by the
one-loop approximation (see \cite{JphysA}), hence the exact results
for $w_{*}$ and $\Omega_{w}$.
The dependence on $\eps$ appears in the {\it corrections}, in particular,
due to the correction exponent $\Omega_{u}$.

\

\noindent 4. $u_{*}=4(\eps-2\xi)/5$, $w_{*}= 2(12\xi-\eps)/15$.
The eigenvalues of the matrix (\ref{OmegaDef}) have the forms:
\begin{eqnarray}
\lambda^{\pm}= \frac{1}{20} \left(11\eps-12 \xi \pm
\sqrt{161\eps^2-824\eps\xi+1104\xi^2} \right).
\label{4npt}
\end{eqnarray}
It is easily checked that they are both real for all $\eps$ and $\xi$
(the expression in the square root is positive definite) and positive
for $\eps/12<\xi<\eps/2$.

This fixed point corresponds to a new nontrivial IR scaling regime
(universality class), in which the nonlinearity of the DP model
(\ref{action}) and the turbulent mixing are simultaneously important;
the corresponding critical exponents depend on the both RG expansion
parameters $\eps$ and $\xi$ and are calculated as double series in these
parameters; see section~\ref{sec:DimeNS}.

\begin{figure}
\begin{center}
\includegraphics[width=11cm]{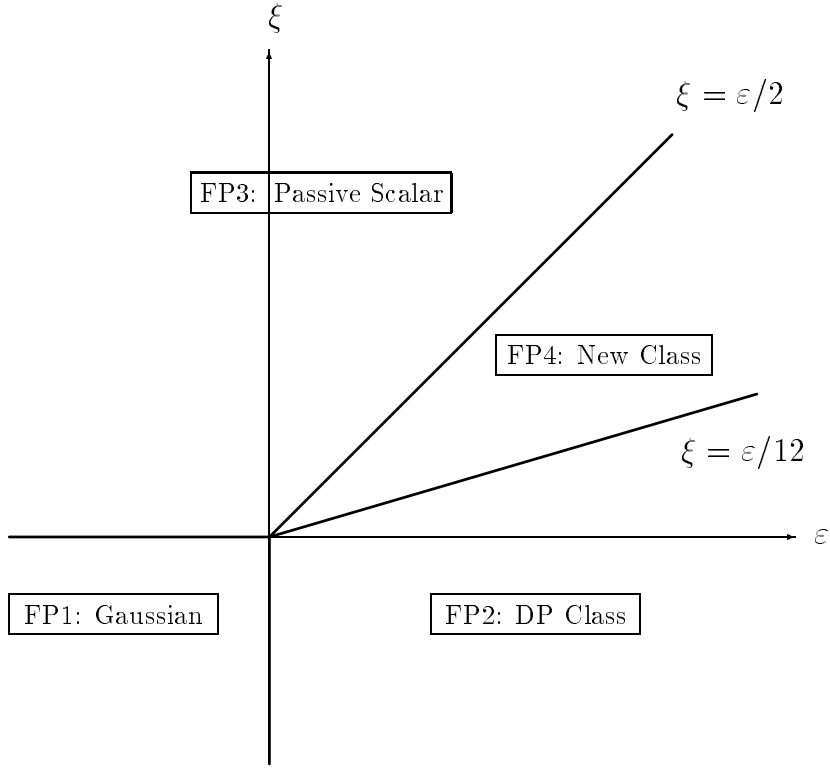}
\caption{\label{fig:pattern}
Regions of stability of
the fixed points in the model (\protect\ref{Action}).}
\end{center}
\end{figure}

In figure~\ref{fig:pattern} we show the regions of IR stability
for all the fixed points in the $\eps$--$\xi$ plane, that is, the regions
in which the eigenvalues of the matrix (\ref{OmegaDef}) for the
given fixed point are both positive.

In the one-loop approximation (\ref{betas2}), all the boundaries of the
regions of stability are given by straight rays; there are neither gaps
nor overlaps between the different regions. \footnote{For the first three
fixed point this is obvious from the expressions for $\Omega_{u,w}$; for
the point 4 this is quite unexpected at the first sight, but can be
explained by the homogeneity of the expressions (\ref{4npt}) in $\eps$
and $\xi$, and is easily seen from the simple form ot the determinant,
${\rm det}\,\Omega = (\eps-2\xi)(12\xi-\eps)/10$.}
Such pattern is typical of the first-order approximations;
cf. \cite{Levy,Beate,AHH,Alexa,Grad6} for various models.
The boundaries $\eps<0$, $\xi<0$ for point 1, $\eps>0$ for point 2
and $\xi>0$ for point 3 are exact, while the other can be affected by
the higher-order corrections: the boundaries will become curved and gaps or
overlaps can appear between the different regions of IR stability.

It is important that, for all these fixed points, the coordinates
$u_{*}$, $w_{*}$ are non-negative in the regions of their IR stability,
in agreement with the physical meaning of the these parameters.
It is also worth noting that the both boundaries for point 4 are determined
by the same eigenvalue $\lambda^{-}$, which changes its sign at
$\xi=\eps/2$ and $\xi=\eps/12$, while $\lambda^{+}$ remains strictly
positive in the entire region of stability.

\section{Critical scaling and critical dimensions} \label{sec:DimeNS}

Existence of an IR attractive fixed point implies existence of scaling
(self-similar) behaviour of the Green functions in the IR range.
In this ``critical scaling'' all the ``IR irrelevant'' parameters
($\lambda$, $\mu$, $g$  and $w$ in our case) are fixed and
the ``IR relevant'' parameters (coordinates/momenta, times/frequencies,
$\tau$ and the fields) are dilated. The critical dimensions $\Delta_{F}$
of the IR relevant quantities $F$ are given by the relations
\begin{eqnarray}
\Delta_{F} = d^{k}_{F}+ \Delta_{\omega} d^{\omega}_{F} + \gamma_{F}^{*},
\qquad  \Delta_{\omega}=2 -\gamma_{\lambda}^{*},
\label{dim}
\end{eqnarray}
with the normalization condition $\Delta_{k} = 1$; see e.g. \cite{Book3}
for more detail. Here $d^{k,\omega}_{F}$ are the canonical dimensions of
$F$, given in table~\ref{table1}, and $\gamma_{F}^{*}$ is the value of
the anomalous dimension (\ref{RGF1}) at the fixed point:
$\gamma_{F}^{*} = \gamma_{F} (u_{*},w_{*})$. This gives
\begin{eqnarray}
\Delta_{\psi} = \Delta_{\psi^{\dag}} = d/2+ \gamma_{\psi}^{*}, \quad
\Delta_{\tau} = 2 + \gamma_{\tau}^{*}.
\label{dimD}
\end{eqnarray}

Four fixed points of the model (\ref{action}) revealed in the preceding
section correspond to four possible IR scaling regimes; for given $\eps$
and $\xi$ only one of them is IR attractive and governs the IR behaviour.
From the general expressions (\ref{dim}), (\ref{dimD})
and the explicit one-loop expressions (\ref{gammasfields}) we find:

\

\noindent 1. Gaussian (free) fixed point; all the expressions are exact:
\begin{eqnarray}
\Delta_{\psi} = d/2, \quad \Delta_{\tau} =  \Delta_{\omega} =  2 .
\label{delta1}
\end{eqnarray}

\

\noindent 2. Directed percolation (DP) regime; mixing irrelevant:
\begin{eqnarray}
\Delta_{\psi} = 2 -7\eps/12, \quad
\Delta_{\tau} = 2- \eps/4, \quad \Delta_{\omega} = 2 - \eps/12 .
\label{delta2}
\end{eqnarray}
These dimensions depend only on $\eps$, the corrections of order
$O(\eps^{2})$ and the references can be found in \cite{JT}.
\footnote{The conventional critical exponents used e.g. in \cite{JT}
are related to the critical dimensions from (\ref{delta2}) as
$z=\Delta_{\omega}$, $1/\nu=\Delta_{\tau}$, $d+\eta=2\Delta_{\psi}$.}
The $O(\eps^{3})$ calculation is in progress; see \cite{AIK}.

\

\noindent 3. Obukhov--Kraichnan exactly soluble regime; all results exact
(see e.g. \cite{JphysA} for the detailed discussion):
\begin{eqnarray}
\Delta_{\omega} = \Delta_{\tau} =  2-\xi, \quad \Delta_{\psi} = d/2.
\label{delta3}
\end{eqnarray}

\

\noindent 4. New universality class (both mixing and DP interaction are
relevant):
\begin{eqnarray}
\Delta_{\psi} = 2 + (\xi -3\eps)/5, \
\Delta_{\tau} = 2- (\eps+ 3\xi)/5, \
\Delta_{\omega} =  2-\xi \ {\rm (exact)} .
\label{delta4}
\end{eqnarray}
The first two dimensions have nontrivial higher-order corrections in $\eps$
and $\xi$. The exact results for $\Delta_{\omega}$ in (\ref{delta3}) and
(\ref{delta4}) follow from the general relations
$\gamma_{\lambda}=\gamma_{w}$ in (\ref{gadf}) and
$\Delta_{\omega}=2-\gamma_{\lambda}^{*}$ in (\ref{dim}) and the identity
$\gamma_{w}= \xi$, which is a consequence of the fixed-point equation
$\beta_{w}=0$ with $\beta_{w}$ from (\ref{betagw})
for any fixed point with $w_{*}\ne 0$.

Let us illustrate the consequences of these general scaling relations for the
spreading of a cloud of the agent (or a cloud of ``infected'' particles) in
the turbulent environment. The mean-square radius $R(t)$ at time $t>0$
of a cloud of such particles, which started from the origin ${\bf x'} = 0$
at time $t'=0$, is related with the linear response function in the
time-coordinate representation as follows:
\begin{eqnarray}
R^{2}(t) = \int d{\bf x}\ x^{2}\, G(t,{\bf x}), \quad
G(t,{\bf x}) = \langle \psi (t,{\bf x}) \psi^{\dag} (0,{\bf 0}) \rangle,
\quad x=|{\bf x}|,
\label{Rad}
\end{eqnarray}
see e.g. \cite{Monin}. For the response function, the relations
(\ref{delta1})--(\ref{delta4}) result in the following IR asymptotic
expression:
\begin{eqnarray}
G(t,{\bf x}) = x^{-2\Delta_{\psi}} \, F
\left(\, \frac{x} { t^ {1/\Delta_{\omega}} }, \,
\frac{\tau}{t^{\Delta_{\tau}/\Delta_{\omega}}}  \right),
\label{Green}
\end{eqnarray}
with some scaling function $F$. Substituting (\ref{Green}) into (\ref{Rad})
gives (with the assumption that the integral converges) the desired
scaling expression for the radius:
\begin{eqnarray}
R^2(t) = t^{ (d+2 -2\Delta_{\psi})/\Delta_{\omega} } \,
f \left(
\frac{\tau}{t^{\Delta_{\tau}/\Delta_{\omega}}}  \right)
\label{R3}
\end{eqnarray}
where the scaling function $f$ is related with $F$ from (\ref{Green}) as
follows:
\[ f (z) = \int d{\bf x}\, x^{2-2\Delta_{\psi}} \, F(x,z). \]
Directly at the critical point (assuming that the function $f$ is finite
at $\tau=0$) one obtains from (\ref{R3}) the power law for the radius:
\begin{eqnarray}
R^2(t) \propto t^{ (d+2 -2\Delta_{\psi})/\Delta_{\omega} } =
t^{(2-2\gamma_{\psi}^{*})/(2-\gamma_{\lambda}^{*})};
\label{R4}
\end{eqnarray}
in the second equality we used the relations (\ref{dim}) and (\ref{dimD}).
For the Gaussian fixed point (\ref{delta1}) the usual ``1/2 law''
$R(t)\propto t^{1/2}$ for the ordinary random walk is recovered. For the
fixed point (\ref{delta3}), where the DP nonlinearity is irrelevant,
one obtains the exact result $R(t)\propto t^{1/(2-\xi)}$. For the most
realistic (Kolmogorov) value $\xi=4/3$ this gives $R(t)\propto t^{3/2}$
in agreement with Richardson's ``4/3 law'' $dR^{2}/dt \propto R^{4/3}$ for
a passively advected scalar impurity; see \cite{Monin}. For the other two
fixed points the exponents in (\ref{R3}), (\ref{R4}) are given by infinite
series in $\eps$ (point 2) or $\eps$ and $\xi$ (point 4); the first-order
approximations are easily obtained from (\ref{delta2}) and (\ref{delta4}).

\section{Conclusion} \label{sec:Conc}

We studied effects of turbulent stirring and mixing on the reaction-diffusion
system in which the spreading of an agent (e.g., infectious disease) occurs,
as an example of critical behaviour in an nonequilibrium system near its
transition between the absorbing and fluctuating phases. We coupled two
paradigmatic models: the so-called simple epidemic process, also known as
Gribov's process and equivalent to Reggeon field theory (\ref{action}),
directed bond percolation process, Schl\"{o}gl's first reaction and so on
\cite{Hinr}--\cite{JT}, and the Obukhov--Kraichnan's rapid-change model
(\ref{Kraich}) for the advecting turbulent velocity field \cite{FGV}.
The full problem can be reformulated as a multiplicatively renormalizable
field theoretic model (\ref{Action}), which allows one to apply the field
theoretic renormalization group \cite{Zinn,Book3} to the analysis of its
critical behaviour.

We showed that, depending on the relation between the spatial dimension
$d$ and the exponent $\xi$ that comes from the Obukhov--Kraichnan's
ensemble, the model exhibits four different critical regimes, associated
with four fixed points of the RG equations.
Three fixed points correspond to known regimes: Gaussian fixed point
(ordinary diffusion or random walk), directed percolation process with
no advection (DP class), and passively advected scalar field (reaction
processes, described by Gribov's nonlinearity, appear irrelevant). The
fourth point reveals existence of a new nonequilibrium universality class,
in which both the reaction and the turbulent mixing are relevant; the
corresponding critical exponents are calculated to the leading order
(one-loop approximation) of the double expansion in $\xi$ and $\eps=4-d$.

Judging naively from the dimensions of the coupling constants
(\ref{charges}) one could expect that the latter regime must take place
when $\xi$ and $\eps$ are simultaneously positive, but the careful RG
analysis shows that the
region of IR stability of the corresponding fixed point is much narrower
(in the one-loop level it shrinks to the sector $\eps/12<\xi<\eps/2$).
Of course, this is not a big surprise: similar phenomenon was established
a long ago for the interplay between the long-range and short-range
spin-spin interactions in models of equilibrium critical behaviour
\cite{Sak}. Later it was also encountered in the long-range versions
of various models with nonequilibrium transitions \cite{Levy,NegFis}
and in models of a passive scalar advected by the velocity field with
finite correlation time \cite{Grad6}.

In our case, this effect leads to interesting prediction: in contrary to
what could be naively anticipated, the most realistic spatial dimensions
$d=2$ or 3 and the Kolmogorov's exponent $\xi=4/3$ for the fully developed
turbulence correspond not to the most nontrivial new regime, but to the
passive-scalar fixed point: the nonlinearity of the Reggeon model is
irrelevant and the spreading of the agent is completely determined by the
turbulent transfer. In particular, the time spreading or a cloud of
infected particles (or of the agent) behaves accordingly to the power law
$R(t)\propto t^{1/(2-\xi)}$, which is the proper generalization of
Richardson's law to the case of arbitrary exponent $\xi$ in the
velocity correlator (\ref{Kraich}).

Further investigation should take into account anisotropy of the
experimental set-up, compressibility, non-Gaussian character and finite
correlation time of the advecting velocity field, effects of memory
(immunization); interaction of the order parameter with other relevant
degrees of freedom (mode-mode coupling), feedback of the reaction on the
dynamics of the velocity and so on. This work is already in progress.

\section*{Acknowledgments}
The authors thank Loran Adzhemyan, Michal Hnatich, Juha Honkonen and
Mikhail Nalimov for helpful discussions. One of the authors (N\,V\,A)
thanks H-K Janssen for bringing his attention to Refs.~\cite{Beate,NegFis}.
The work was supported in part by the Russian Foundation for Fundamental
Research (grant No~08-02-00125a), the Russian National Program
(grant No~2.1.1.1112) and the program ``Russian Scientific Schools''
(grant No~5538.2006.2).

\appendix
\section{Calculation of the renormalization constants} \label{sec:Z}

In this section we derive the first-order results (\ref{Zone}) for the
renormalization constants. Of course, the one-loop calculation is rather
simple and can be accomplished in a few different ways, but we will
discuss it for completeness and in order to mention some interesting
subtleties specific of the model (\ref{Action}).

The renormalization constants can be found from the requirement that
the Green functions of the renormalized model (\ref{RenAct}), when
expressed in renormalized variables, be UV finite (in our case, be
finite at $\eps\to0$, $\xi\to0$). Owing to the symmetry (\ref{symm}) and
to the Galilean invariance, the full set of constants $Z_{1}$--$Z_{4}$
in our model can be found from just two 1-irreducible functions:
$\langle \psi^{\dag} \psi \rangle_{\rm 1-ir}$ and
$\langle \psi^{\dag} \psi^{\dag} \psi  \rangle_{\rm 1-ir}$ (or, if desired,
$\langle \psi^{\dag} \psi \psi \rangle_{\rm 1-ir}$).
In the renormalized model, the corresponding one-loop approximations have
the forms
\begin{eqnarray}
\langle \psi^{\dag} \psi \rangle_{\rm 1-ir} = {\rm i} \omega Z_{1}
+ \lambda p^{2}  Z_{2} + \lambda \tau  Z_{3} - \frac{1}{2}\
\raisebox{-0.41cm}{\includegraphics[width=2.3cm]{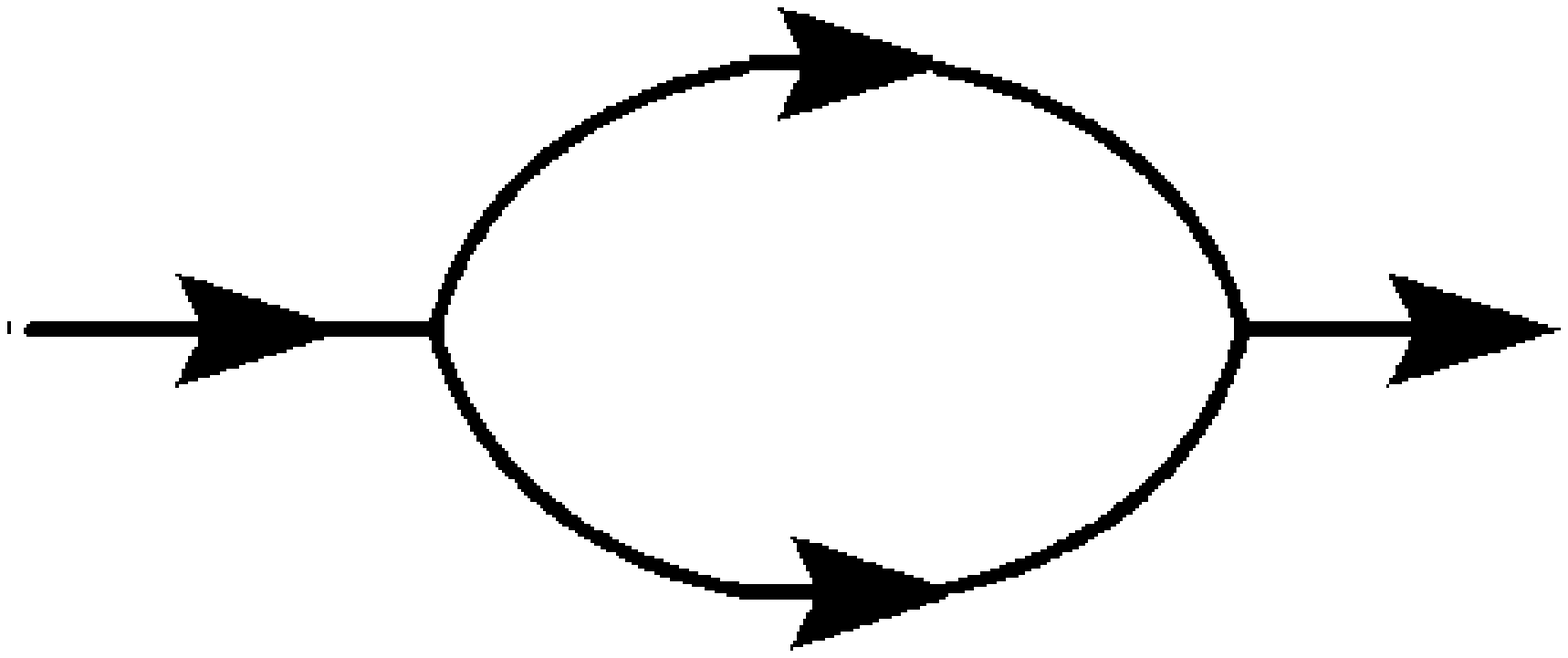}} \ -
\
\raisebox{-0.00cm}{\includegraphics[width=2.3cm] {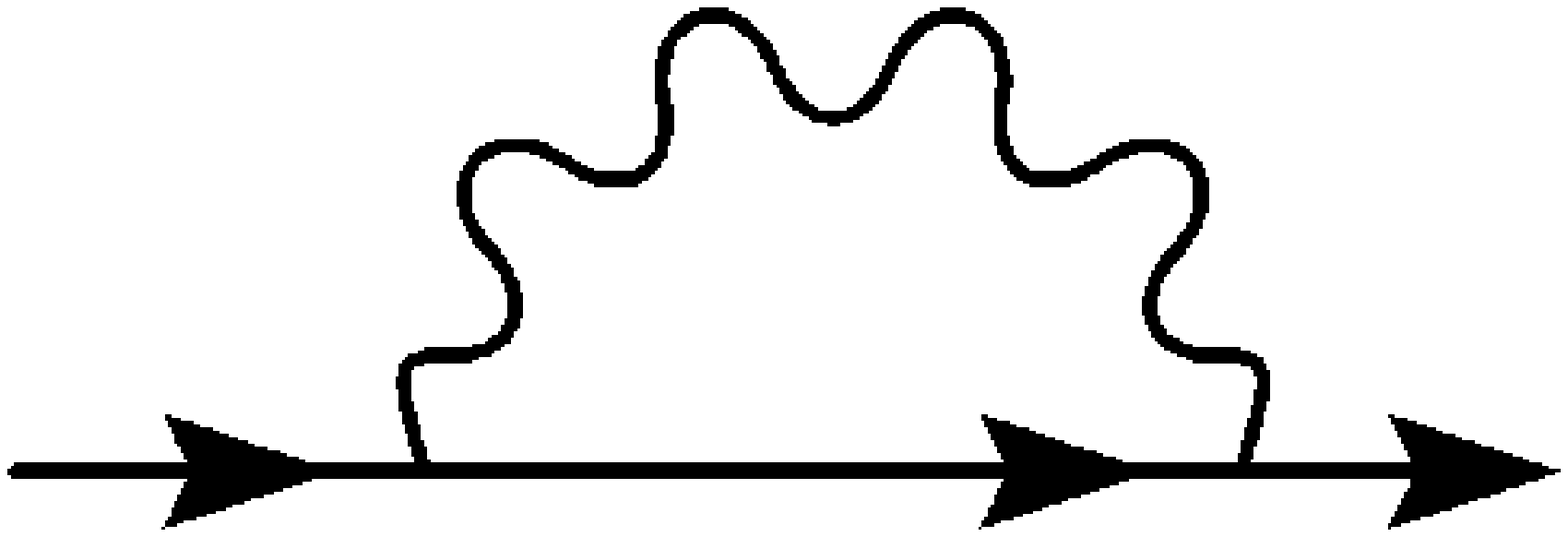}}
\nonumber \\ {} \label{Diagr1}
\end{eqnarray}

(which is the Dyson equation for the exact function
$\langle \psi^{\dag} \psi \rangle$, hence the minus signs in front of
the diagrams) and
\begin{eqnarray}
\langle \psi^{\dag} \psi^{\dag} \psi  \rangle_{\rm 1-ir} = g Z_{4}
+ 2
\raisebox{-0.60cm}{\includegraphics[width=1.7cm]{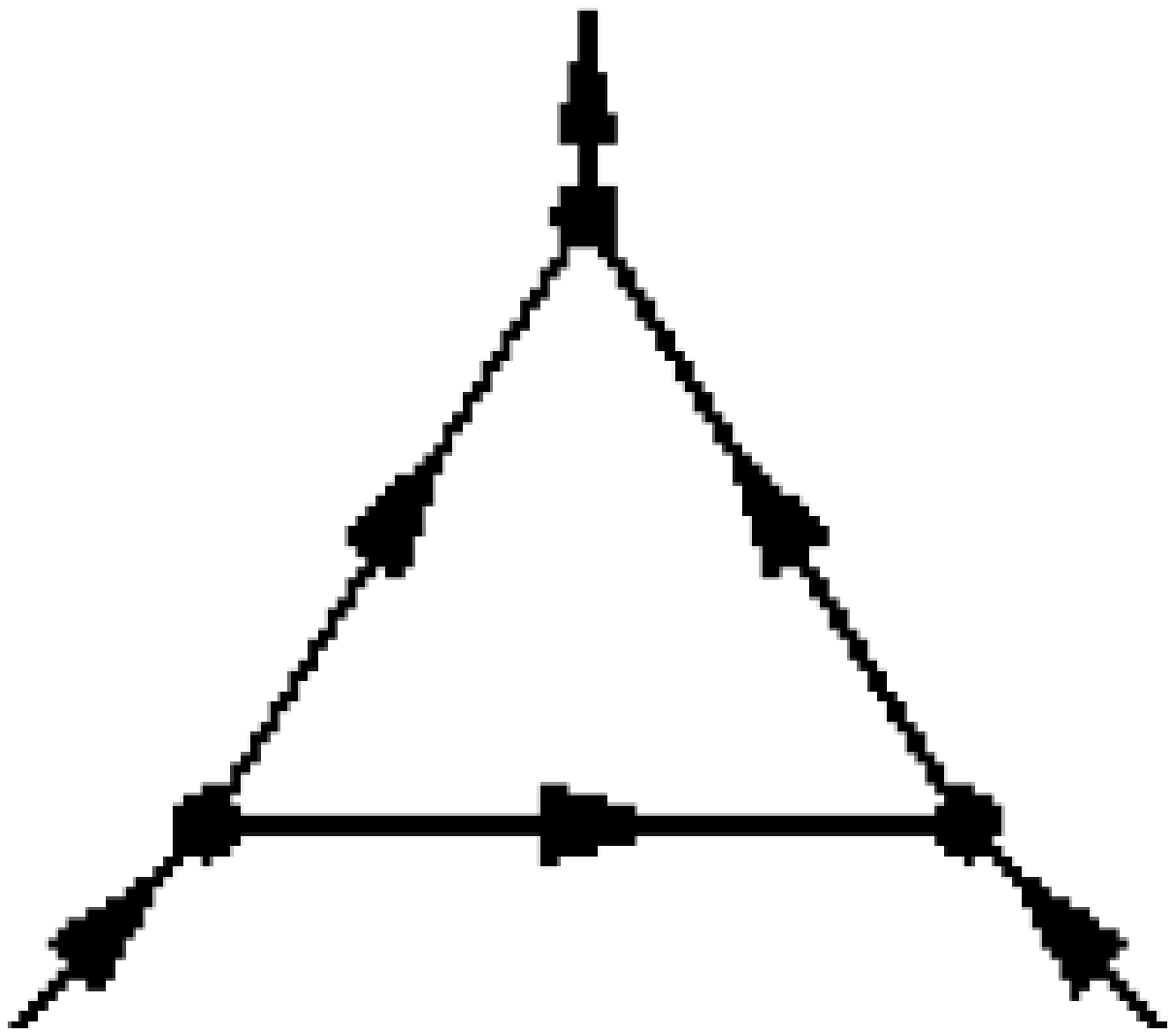}}
+ 2 \raisebox{-0.60cm}{\includegraphics[width=1.65cm]{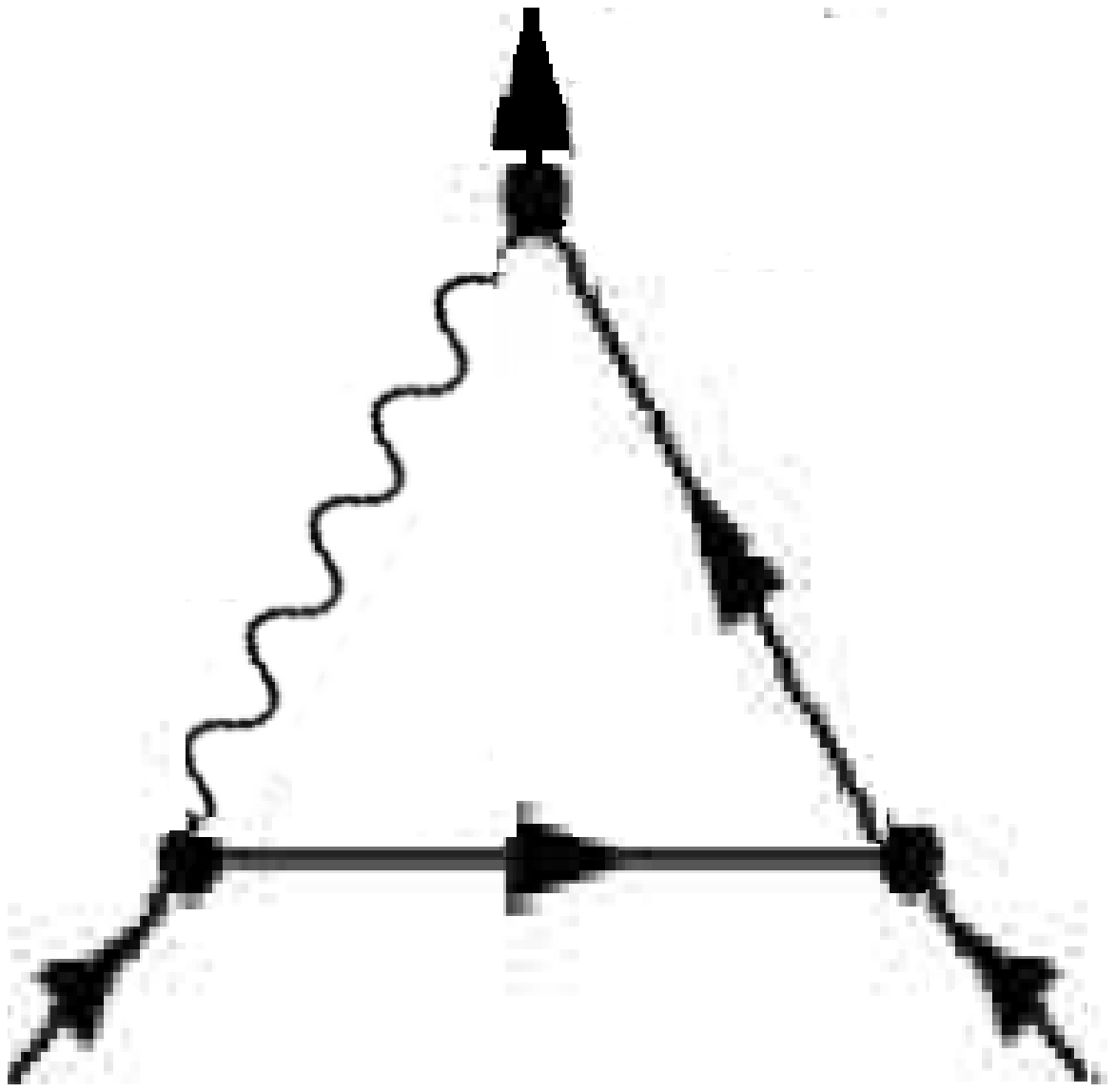}}
+ \frac{1}{2}\,
\raisebox{-0.60cm}{\includegraphics[width=1.7cm]{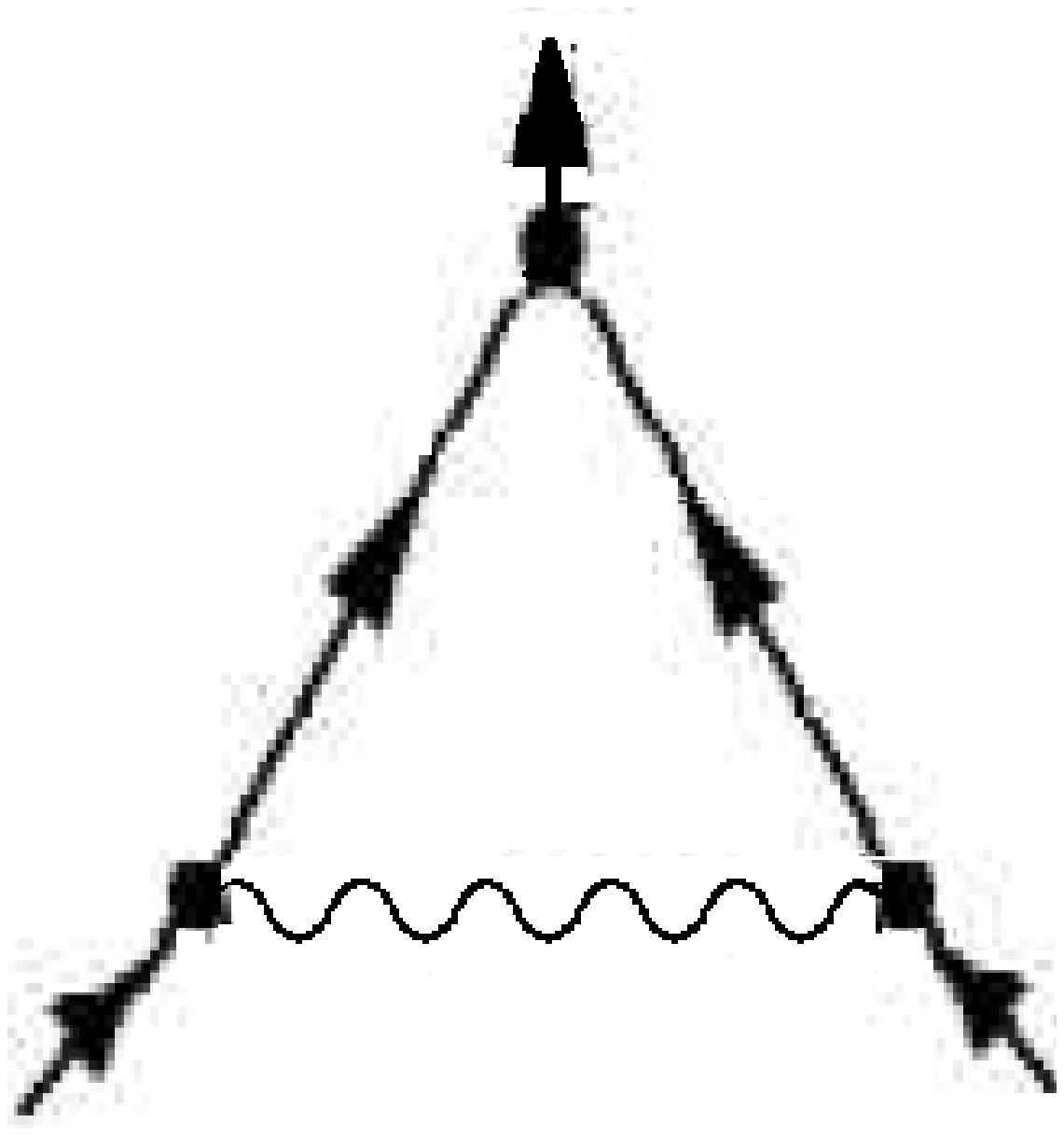}}
%\nonumber
\label{Diagr2}
\end{eqnarray}
The first two diagrams in (\ref{Diagr2}) have in fact two different forms,
related by the mirror reflection, but they give equal contributions to
the renormalization constants and are accounted by the factors of 2.
Owing to the Galilean symmetry, the constant $Z_{1}$ can also be found
from the function
\begin{eqnarray}
\langle \psi^{\dag} \psi {\bf v}  \rangle_{\rm 1-ir} = - {\rm i} {\bf p}
Z_{1} +
\raisebox{-0.60cm}{\includegraphics[width=1.7cm]{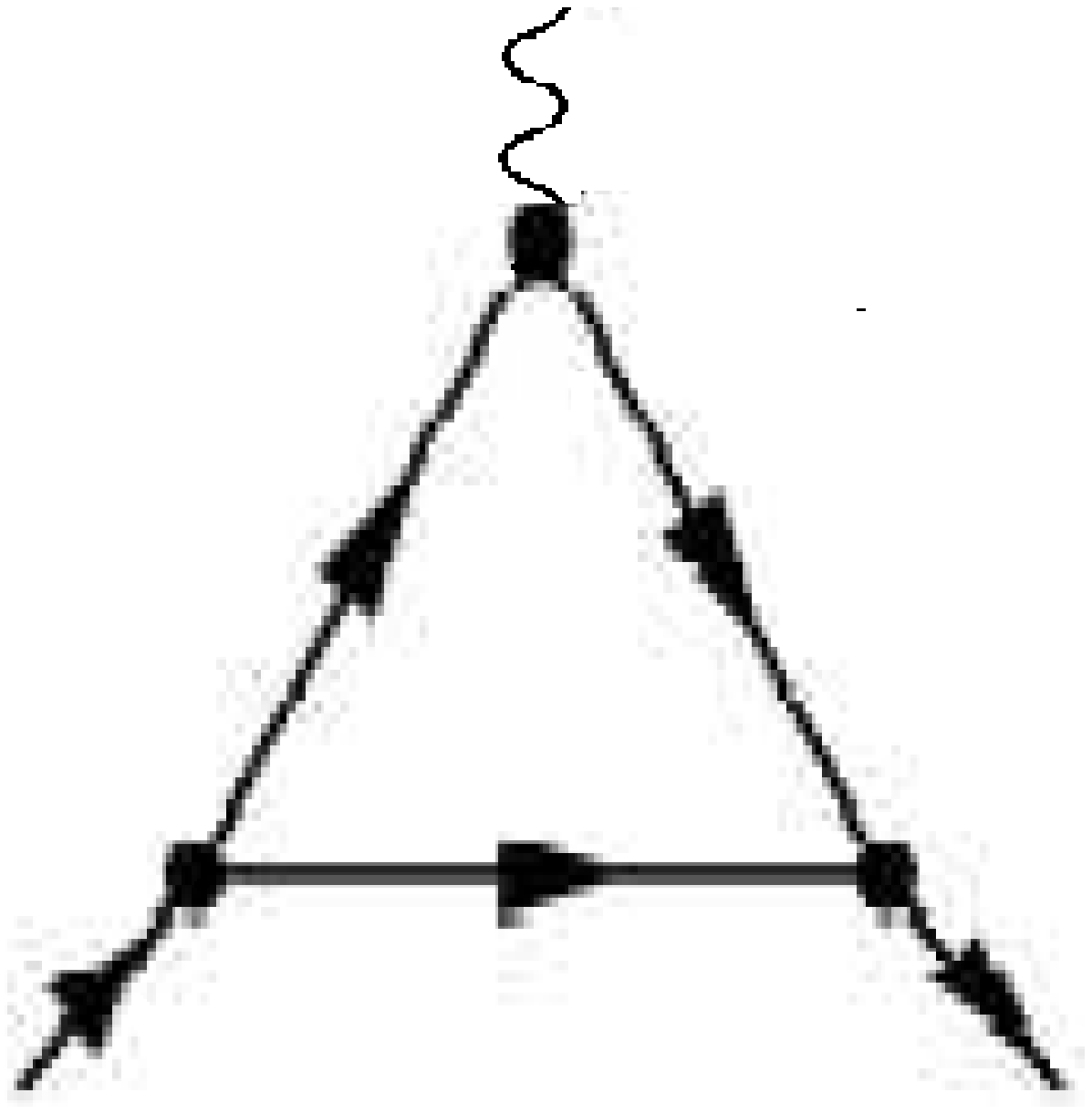}}
\ +\
\raisebox{-0.60cm}{\includegraphics[width=1.7cm]{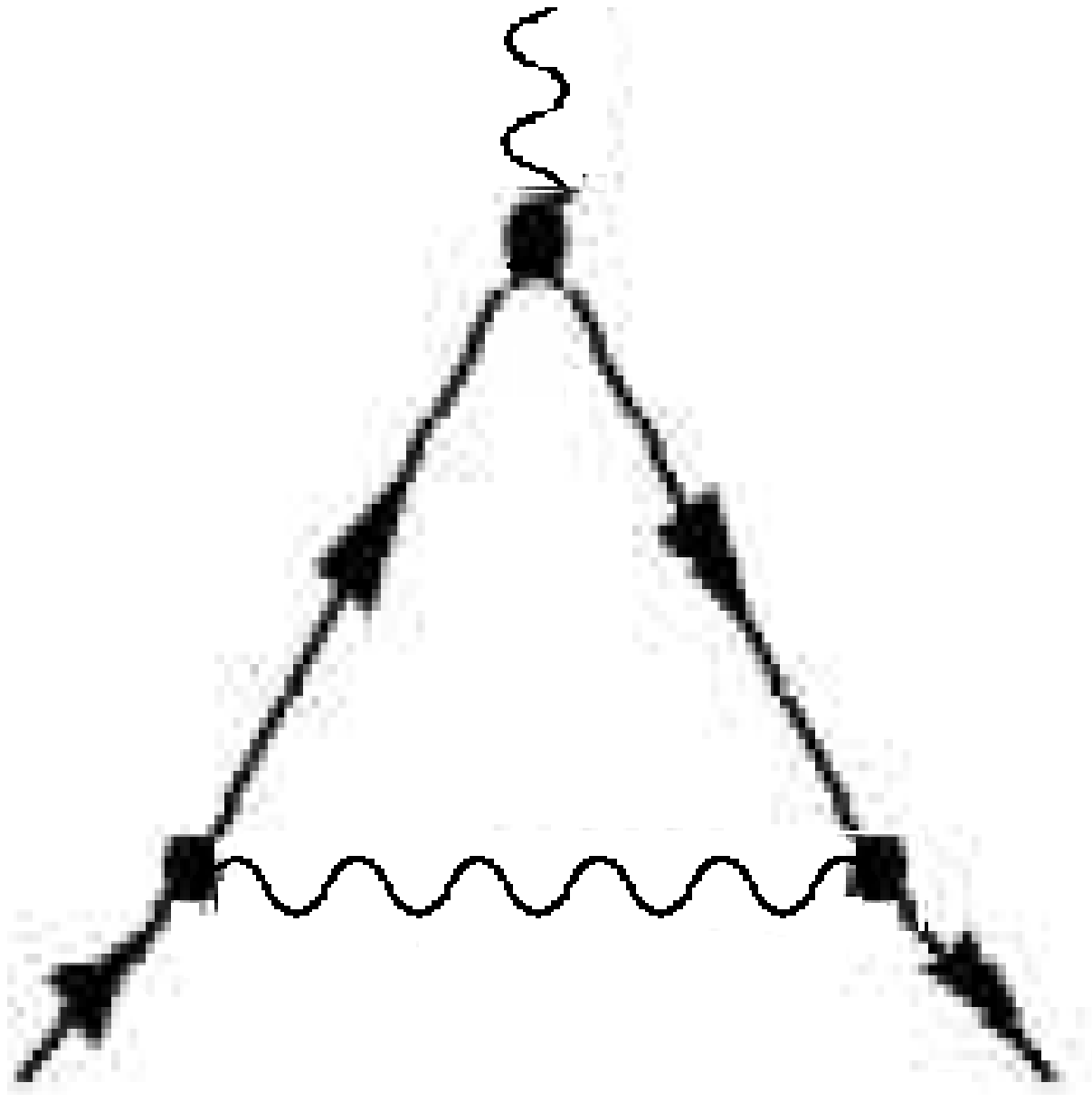}} .
%\nonumber
\label{Diagr3}
\end{eqnarray}
The solid lines in (\ref{Diagr1})--(\ref{Diagr3}) denote the propagator
$\langle \psi \psi^{\dag} \rangle_{0}$ from (\ref{lines})
(the arrow is directed from $\psi$ to $\psi^{\dag}$) and the wavy lines
correspond to $\langle vv \rangle_{0}$ from (\ref{Kraich}).
All the diagrammatic elements should be expressed in renormalized variables
using the relations (\ref{RenAct})--(\ref{ZZ}).
In the one-loop approximation, the $Z$'s in the bare terms of
(\ref{Diagr1}), (\ref{Diagr2}) should be taken in the first order
in $u= g^{2}$ and $w$, while in the diagrams they should simply be replaced
with unities, $Z_{i}\to1$. Thus the passage to renormalized
variables in the diagrams is achieved by the simple substitutions
$\lambda_{0} \to \lambda$, $\tau_{0} \to \tau$,
$g_{0} \to g\mu^{\eps/2}$ and $w_{0} \to w\mu^{\xi}$.

The second and third diagrams in (\ref{Diagr2}) and the second diagram
in (\ref{Diagr3}) appear UV finite and therefore give no contribution
to the renormalization constants. Indeed, due to the transversality of
the velocity field, the derivative in the vertex
$-\psi^{\dag} (v\partial) \psi$ can also be moved onto the field
$\psi^{\dag}$ using integration by parts:
$-\psi^{\dag} (v\partial) \psi = \psi (v\partial) \psi^{\dag}$. Thus
in any diagram involving $n$ external vertices of this type, the factor
$p^{n}$ with $n$ {\it external} momenta $p$ will be taken outside the
corresponding integrals. This reduces the dimension of the integrand by
$n$ units and can make it UV convergent. In the case at hand, this proves
the UV finiteness of the three diagrams mentioned above: for all of them
$n=2$ while the formal index of divergence is 0 or 1.

What is more, since the propagator (\ref{lines}) is retarded and
(\ref{Kraich}) contains the $\delta$ function in time, the second
diagrams both in (\ref{Diagr2}) and (\ref{Diagr3}) contain
self-contracted ``circuits'' of the step functions in time and therefore
vanish identically. This argument, however, does not apply to the second
diagram in (\ref{Diagr1}), which requires a more careful treatment; cf.
\cite{Alexa,JphysA}. The analytic expression for that diagram has the form
\begin{eqnarray}
- p_{i}p_{j} \int \frac{d\omega}{(2\pi)} \int_{k>m}
\frac{d{\bf k}}{(2\pi)^{d}} \,
 \frac{D_{0}P_{ij}({\bf k})}{k^{d+\xi}}\,
\frac{1} {  -{\rm i} \omega + \sigma(\p-{\bf k}) },
\label{D2}
\end{eqnarray}
where the prefactor comes from the vertices, the IR cutoff and the first
cofactor in the integrand come from the propagator (\ref{Kraich}) and the
second cofactor with $\sigma({\bf k})= \lambda(  k^{2}+\tau)$ is the
propagator (\ref{lines}). The expression (\ref{D2}) is independent
of the external frequency. Integration over $\omega$ involves the
indeterminacy
\begin{eqnarray}
\int \frac{d\omega}{(2\pi)} \, \frac{1}{-{\rm i} \omega +
\sigma(\p-{\bf k})} =  \theta(0),
\label{inde}
\end{eqnarray}
where $\theta(0)$ is the step function at the origin. This reflects the
details of the velocity statistics lost in the white-noise limit; see the
discussion in \cite{FGV}. In the case at hand, the $\delta$ function in
(\ref{white}) should be understood as the limit of a narrow function which
is necessarily {\it symmetric} in $t$, $t'$, because one deals with a pair
correlator. Thus the indeterminacy in (\ref{inde}) must be unambiguously
resolved as half the sum of the limits: $\theta(0)=1/2$. Then the remaining
integral over ${\bf k}$ in (\ref{D2}) appears independent of ${\bf p}$ and
$\tau$ and is easily calculated:
\begin{eqnarray}
\int_{k>m} \frac{d{\bf k}}{(2\pi)^{d}} \, \frac{P_{ij}({\bf k})}{k^{d+\xi}}
= \frac{\delta_{ij}\,(d-1)}{d}\, \int_{k>m} \frac{d{\bf k}}{(2\pi)^{d}} \,
\frac{1}{k^{d+\xi}} =
\delta_{ij} m^{-\xi} \frac{S_d}{(2\pi)^{d}} \frac{(d-1)}{d \xi},
\nonumber
\end{eqnarray}
where $S_d=2\pi^{d/2}/\Gamma(d/2)$ with Euler's $\Gamma$ function is the
surface area of the unit sphere in the $d$-dimensional space. Collecting all
the factors and setting $D_{0} = \lambda_{0} w_{0} = \lambda w \mu^{\xi}$
gives the final result for the diagram:
\begin{eqnarray}
- \lambda w p^{2} (\mu/m)^{\xi}\, \frac{(d-1)}{2d \xi}\frac{S_d}{(2\pi)^{d}}
= - \lambda p^{2} \frac{w}{16\pi^2} \frac{3}{4\xi} + {\rm UV\ finite\ part}.
\label{34}
\end{eqnarray}
In the last equality in (\ref{34}), the UV divergent part of the diagram is
selected (the replacements $(\mu/m)^{\xi}\to1$ and $d=4-\eps\to4$ are made);
it contains a first-order pole in $\xi$. The expression (\ref{34}) as a whole
is proportional to $p^{2}$, so that it gives a contribution only to the
renormalization constant $Z_{2}$ in (\ref{Diagr1}).

The remaining two diagrams in (\ref{Diagr1}), (\ref{Diagr2}) do not involve
the velocity correlator; they are independent of $\xi$ and contain only
poles in $\eps$. Although the calculation of these diagrams is discussed
in \cite{JT} within the context of the model (\ref{action}), we will
sketch an alternative calculation here, mainly in order to present a
reference formula, which can be interesting in itself.

The key point is as follows: the convolution of two functions of the form
\begin{equation}
F(\alpha,a,\tau) \equiv (-{\rm i}\omega\, a + k^{2} +\tau)^{-\alpha}, \quad
\tau>0
\label{fuNk}
\end{equation}
is a function of the same form:
\begin{equation}
F(\alpha_{1},a_{1},\tau_{1}) * F(\alpha_{2},a_{2},\tau_{2})
=  K (\alpha_{1},\alpha_{2}; a_{1},a_{2})\,
F(\alpha_{3}, a_{3}, \tau_{3})
\label{conv1}
\end{equation}
if $a_{1}$ and $a_{2}$ have the same sign and zero otherwise. Here
\begin{equation}
a_{3}=a_{1}+a_{2}, \quad \alpha_{3}=\alpha_{1}+\alpha_{2}-d/2-1, \quad
\tau_{3} = a_{3} (\tau_{1}/a_{1}+\tau_{2}/a_{2});
\label{comt}
\end{equation}
the coefficient is independent on the $\tau_{1,2}$ and has the form:
\[ K (\alpha_{1},\alpha_{2}; a_{1},a_{2})
= a_{1}^{d/2-\alpha_{1}} a_{2}^{d/2-\alpha_{2}}
a_{3}^{\alpha_{1}+\alpha_{2}-d-1} \, \frac{\Gamma(\alpha_{3})}
{(4\pi)^{d/2} \Gamma(\alpha_{1}) \Gamma(\alpha_{2})}.  \]

For the ``massless'' case $\tau_{1,2}=0$ this formula was proposed and used
in the three-loop calculation of the critical exponent $z$ in the so-called
model A of critical dynamics in \cite{AV}; see also \cite{Alexa,AH} for
other applications. For the general case ($\tau_{1,2}\ge 0$) equation
(\ref{conv1}) can be obtained from the following observation:
In the time-space representation the function (\ref{fuNk}) takes on the form
\begin{equation}
F(\alpha,a,\tau) \to \frac{\theta(t\,{\rm sign} (a))\, a^{d/2-\alpha}}
{(4\pi)^{d/2}\, \Gamma(\alpha)}\, t^{\alpha-d/2-1} \,
\exp \left\{ - \frac{ax^{2}}{4t} -\tau t/a  \right\},
\label{xt}
\end{equation}
and the product of such functions (which corresponds to the convolution
of their Fourier transforms) is obviously a function of the same form.
Note that, when $a_{1}$ and $a_{2}$ have different signs, the convolution
in (\ref{conv1}) corresponds to the product of a retarded and an
advanced functions of the form (\ref{xt}) and therefore vanishes.

Let us turn to the remaining diagrams in (\ref{Diagr1}), (\ref{Diagr2}).
With no loss of generality, one can set $\lambda=1$ (the dependence on
$\lambda$ can be restored in the final answers using the dimensionality
considerations). Then the frequency-momentum integral that corresponds to the
first diagram in (\ref{Diagr1}), is equal to the convolution (\ref{conv1}),
in which $a_{1,2}=1$, $\alpha_{1,2}=1$, $\tau_{1,2}=\tau$. From (\ref{comt})
one obtains $a_{3}=2$, $\alpha_{3}=-1+\eps/2$, $\tau_{3}=4\tau$,
so that the left-hand side of (\ref{conv1}) is proportional to
\begin{eqnarray}
\Gamma(-1+\eps/2) (-2{\rm i}\omega+k^{2}+4\tau)^{1-\eps/2} =
\frac{2}{\eps} (-2{\rm i}\omega+k^{2}+4\tau) + {\rm UV\, finite\ part}
\nonumber \\ {}
\label{D1}
\end{eqnarray}
(here and below we omit uninteresting finite numerical factors, powers
of $\pi$ and so on).

The first diagram in (\ref{Diagr2}) is logarithmically divergent.
According to the general statements of the renormalization theory,
its divergent part in the MS scheme does not depend on the specific
choice of the external momenta, frequencies, and ``masses''  like $\tau$,
provided this choice guarantees IR convergence of the corresponding
integral; see e.g. \cite{Zinn,Book3}. Thus we can set the external
frequency and momentum flowing into the right lower vertex equal to zero.
(We could also set $\tau=0$, but this is not necessary).
Then the integral that to the diagram in the frequency-momentum
representation, becomes equal to the convolution (\ref{conv1})
with $a_{1,2}=1$, $\alpha_{1}=1$, $\alpha_{2}=2$, $\tau_{1,2}=\tau$.
From (\ref{comt}) it follows $a_{3}=2$, $\alpha_{3}=\eps/2$,
$\tau_{3}=4\tau$, and the left-hand side of (\ref{conv1}) is proportional to
\begin{eqnarray}
\Gamma(\eps/2) (-2{\rm i}\omega+k^{2}+4\tau)^{-\eps/2} =
\frac{2}{\eps}  + {\rm UV\, finite\ part}.
\label{Di2}
\end{eqnarray}

As expected, the expressions (\ref{D1}) and (\ref{Di2}) contain first-order
poles in $\eps$. Their pole parts are polynomials in frequencies, momenta,
and ``masses,'' so that they can be cancelled in expressions (\ref{Diagr1}),
(\ref{Diagr2}) by the proper choice of the renormalization constants
$Z_{1}$--$Z_{4}$. Taking into account all the factors (signs, symmetry
coefficients, factors $\pm g$ from the vertices) and the replacement
(\ref{16p}) gives the results announced in
(\ref{Zone}). It remains to note that the calculation of the 1-irreducible
function (\ref{Diagr3}) indeed gives the same result (\ref{Zone}) for the
constant $Z_{1}$, in agreement with general consequences of the Galilean
invariance of our model.

\end{document}